\documentclass[]{article}

\usepackage{amsmath, amssymb, amsthm, bbm, booktabs, color, enumitem, graphicx, multicol, multirow, url}

\usepackage[a4paper, total={6.25in, 9.0in}]{geometry}
\usepackage[authoryear]{natbib}
\usepackage[colorlinks,citecolor=blue]{hyperref}
\usepackage[labelfont=bf]{caption}

\newcommand{\hsp}{\hspace{0.2mm}}
\newcommand{\myE}{\mathbb{E}}
\newcommand{\one}{\mathbbm{1}}

\newcommand{\CRPS}{\operatorname{CRPS}}
\newcommand{\PC}{\operatorname{PC}}
\newcommand{\PCS}{\operatorname{PCS}}

\begin{document}

\title{Probabilistic measures afford fair comparisons \\ of AIWP and NWP model output} 
\author{Tilmann Gneiting$^{1,2}$, Tobias Biegert$^2$, Kristof Kraus$^{1,3}$, Eva-Maria Walz$^{1,2}$, \\ Alexander I.~Jordan$^{1,2}$, Sebastian Lerch$^{4,1,2}$ \\
\vspace{0.5cm} \\ 
\small $^1$Computational Statistics group, Heidelberg Institute for Theoretical Studies, Heidelberg, Germany \\
\small $^2$Institute of Statistics, Karlsruhe Institute of Technology (KIT), Karlsruhe, Germany \\
\small $^3$Institute for Stochastics, Karlsruhe Institute of Technology (KIT), Karlsruhe, Germany \\
\small $^4$Department of Mathematics and Computer Science, University of Marburg, Marburg, Germany}
\maketitle

\begin{abstract}

We introduce a new measure for fair and meaningful comparisons of single-valued output from artificial intelligence based weather prediction (AIWP) and numerical weather prediction (NWP) models, called potential continuous ranked probability score ($\PC$).  In a nutshell, we subject the deterministic backbone of physics-based and data-driven models post hoc to the same statistical postprocessing technique, namely, isotonic distributional regression (IDR).  Then we find $\PC$ as the mean continuous ranked probability score (CRPS) of the postprocessed probabilistic forecasts.  The nonnegative $\PC$ measure quantifies potential predictive performance and is invariant under strictly increasing transformations of the model output.  $\PC$ attains its most desirable value of zero if, and only if, the weather outcome $Y$ is a fixed, non-decreasing function of the model output $X$.  The $\PC$ measure is recorded in the unit of the outcome, has an upper bound of one half times the mean absolute difference between outcomes, and serves as a proxy for the mean CRPS of real-time, operational probabilistic products.  When applied to WeatherBench 2 data, our approach demonstrates that the data-driven GraphCast model outperforms the leading, physics-based European Centre for Medium Range Weather Forecasts (ECMWF) high-resolution (HRES) model.  Furthermore, the $\PC$ measure for the HRES model aligns exceptionally well with the mean CRPS of the operational ECMWF ensemble.  Across application domains, our approach affords comparisons of single-valued forecasts in settings where the pre-specification of a loss function --- which is the usual, and principally superior, procedure in forecast contests, administrative, and benchmarks settings --- places competitors on unequal footings.

\smallskip
\textit{Keywords}: Easy Uncertainty Quantification (EasyUQ), GraphCast, Integrated Forecast System (IFS), Pangu-Weather, verification, WeatherBench 
\end{abstract}

\section{Introduction}  \label{sec:introduction}

In a fast-breaking, revolutionary recent development, deterministic forecasts from data-driven artificial intelligence (AI) based weather prediction (AIWP) models now appear to outperform forecasts from classical physics-based numerical weather prediction (NWP) models \citep{BenBouallegue2024, Kochkov2024, Bodnar2025}.  Nonetheless, the superior predictive ability of AIWP forecasts might still get questioned, due to an ongoing debate over fair ways of comparing AIWP and NWP forecasts, particularly in benchmarks settings \citep{Rasp2020, Rasp2024, Radford2025a}. 

To provide background on the evaluation of deterministic forecasts, \citet{Murphy1985} and \citet{Gneiting2011a} have argued persuasively that forecast models need directives in the form of either a pre-specified loss function (e.g., squared error) or a pre-specified functional (e.g., the mean).  Accordingly, root mean squared error (RMSE) has been a preferred measure in comparisons of AIWP and NWP forecasts.  However, comparisons in terms of squared error, or any other pre-specified loss function, might put AIWP at an unfair advantage over NWP.  Indeed, AIWP models operate by optimizing training loss --- see \citet{Brenowitz2025} and \citet{Selz2025} for recent discussions --- and they can use the pre-specified loss function for this purpose, unlike NWP models, whence the reliance on RMSE might unduly favour AIWP.  For example, \citet[p.~7932]{Olivetti2024} note that ``the RMSE is also the objective function of the machine learning (ML) models, which means that evaluating against RMSE is not a fully independent benchmark.''  Another matter of concern arises as, in key benchmark settings \citep{Rasp2024}, some models are initialized with, and evaluated against, the Integrated Forecast System (IFS) analysis as ground truth, whereas other models are initialized with, and evaluated against, the ERA5 analysis, which raises the question whether direct comparisons are meaningful. 

Furthermore, AIWP models are usually compared to raw NWP forecasts, which is counter to extant practice, in which NWP model output is subject to statistical postprocessing prior to public release \citep{Vannitsem2021}.  These considerations exacerbate the aforementioned concerns that comparisons in the extant literature might favour AIWP models.  While one might follow \citet[p.~23]{Rasp2024} and argue that for a practically relevant and fair comparison, AIWP model output ought to be compared to statistically postprocessed NWP model output, AIWP models might benefit from statistical postprocessing as well.

Recently, \citet[pp.~1 and 9]{Brenowitz2025} noted ``a need for benchmarking deterministic forecasts from the lens of probabilistic skill'' and devised an innovative framework based on lagged initial condition ensembles that ``allows one to compare physics-based and data-driven models with the same ensemble technique''.  Their approach enriches deterministic model output to lagged forecast ensembles, and then takes the mean continuous ranked probability score (CRPS) of the lagged ensemble as a measure of predictive ability.  This addresses the loss function problem in the comparison of AIWP and NWP models in a smart and efficient way, but does not alleviate the aforementioned other matters of concern.  

Our approach is of a similar spirit, as it also converts deterministic AIWP and NWP model output into probabilistic forecasts and then takes the mean CRPS of the thus generated probabilistic forecasts as a measure of predictive ability.  However, we deviate crucially from \citet{Brenowitz2025} in the way that the conversion from deterministic model output to probabilistic forecasts is done.  Specifically, we employ the recently developed Easy Uncertainty Quantification (EasyUQ) method \citep{Walz2024a}, which is a special case of isotonic distributional regression \citep[IDR,][]{Henzi2021}, as a statistical postprocessing technique on the evaluation set, to convert deterministic model output into calibrated predictive distributions.  Unlike the generation of lagged ensembles, our method of conversion is fully automated and does not require tuning nor other types of user intervention, and it employs model output at the prediction horizon under consideration only.  Furthermore, by subjecting the deterministic backbone of AIWP and NWP forecasts to the very same statistical postprocessing technique that is proven to be optimal in terms of CRPS \citep{Henzi2021}, the probabilistic forecasts thus generated adapt to the choice of ground truth, and the two types of forecasts are treated equally with respect to postprocessing. 

The potential CRPS ($\PC$) measure that we introduce here equals the mean CRPS of the post hoc (in-sample) IDR postprocessed predictive distributions.  $\PC$ is negatively oriented --- the smaller, the better --- and attains the most desirable value of zero if, and only if, the model output allows for perfect predictions.  The largest possible value of $\PC$ is one half times the mean absolute difference between the outcomes in the evaluation set.  Like other commonly used measures that quantify potential predictive ability, such as the popular area under the curve (AUC) criterion, $\PC$ is invariant under strictly increasing transformations of the model output.  Nonetheless, $\PC$ is reported in the same unit as the outcome and serves as a proxy for the CRPS of operational probabilistic products.

The remainder of the paper is organized as follows.  The key technical development is in Section \ref{sec:methods}, where we detail the algorithms used to compute the $\PC$ and $\PC$ skill ($\PCS$) measures, describe key properties of these measures, and present a brief simulation study.  In Section \ref{sec:WeatherBench} we apply the $\PC$ and $\PCS$ measures to WeatherBench 1 \citep{Rasp2020} and WeatherBench 2 \citep{Rasp2024} data.  While in WeatherBench 1 AIWP models showed far weaker predictive ability than NWP models, the past few years have seen a dramatic reverse.  The tremendous progress is evidenced by our results for WeatherBench 2, where we compare the data-driven Pangu-Weather \citep{Bi2023} and GraphCast \citep{Lam2023} models to the leading physics-based European Centre for Medium Range Weather Forecasts (ECMWF) high-resolution (HRES) model.  Furthermore, the $\PC$ measure for the HRES model serves as a very close proxy for the mean CRPS of the operational ECMWF ensemble.  We close the paper with a discussion in Section \ref{sec:discussion}.

\section{Methods}  \label{sec:methods}

Our methods rely on two key tools, namely, the continuous ranked probability score (CRPS), which is a popular and widely used loss function for the evaluation of probabilistic forecasts, and the recently proposed Easy Uncertainty Quantification (EasyUQ) method that converts deterministic model output into probability distributions.  

To define the CRPS, we identify a probabilistic forecast with the corresponding cumulative distribution function (CDF), $F$, and let $y$ be the outcome.  Then
\begin{align}  
\CRPS(F,y) 
& = \int_{-\infty}^\infty \left( F(x) - \one \{ \hsp y \leq x \} \right)^2 \mathrm{d} x \label{eq:CRPS1} \\
& = \myE_F \! \left| Y - y \right| - \frac{1}{2} \hsp\hsp \myE_F \! \left| Y - Y' \right| \! , \label{eq:CRPS2} 
\end{align}
where $\myE$ denotes the expectation operator and $Y$ and $Y'$ are independent random variables with distribution $F$ and finite first moment \citep{Gneiting2007a}.

\citet{Walz2024a} devised EasyUQ as a special case of the isotonic distributional regression (IDR) technique, developed by \citet{Henzi2021}, that converts deterministic model output into calibrated probabilistic forecasts.  The starting point is data 
\begin{align}  \label{eq:data}
\left( x_1, y_1 \right), \ldots, \left( x_n, y_n \right), 
\end{align}
where $x_i$ is real-valued model output and $y_i$ is the respective outcome, for $i = 1, \ldots, n$.  The EasyUQ technique then finds probabilistic forecasts $\hat{F}_1, \ldots, \hat{F}_n$ that minimize the target criterion 
\begin{align*}  
\frac{1}{n} \sum_{i = 1}^n \CRPS(F_i,y_i), 
\end{align*} 
subject to the natural constraint of isotonicity with respect to the model output, in the technical sense that if $x_i \leq x_j$ then $q_\alpha(F_i) \leq q_\alpha(F_j)$ for all $\alpha \in (0,1)$, where $q_\alpha$ denotes the $\alpha$-quantile function.  The constrained minimization problem has an analytic solution $\hat{F}_1, \ldots, \hat{F}_n$ in terms of $x_1, \ldots, x_n$ and $y_1, \ldots y_n$, this solution is unique and calibrated, and it is simultaneously optimal under a comprehensive class of proper scoring rules \citep{Henzi2021, Walz2024a}.  In plain words, the EasyUQ solution is the best probabilistic forecast that can be constructed post hoc from the data at \eqref{eq:data}, subject to the natural constraint that the induced quantile forecasts share the order relations of the original model output.   

\subsection{Potential CRPS (PC) measure}  \label{sec:PCS}

We now describe the computation of the potential CRPS ($\PC$) measure, when the goal is the comparative evaluation of deterministic output from $J$ competing models.  Specifically, suppose we are given evaluation data 
\begin{align*}  
\left( x_1^{(1)}, \ldots, x_1^{(J \hsp)}, y_1 \right), \ldots, \left( x_n^{(1)}, \ldots, x_n^{(J \hsp)}, y_n \right),
\end{align*}
where $x_i^{(j)}$ is the forecast from model $j = 1, \ldots, J$, and $y_i$ is the respective outcome, for evaluation instance $i = 1, \ldots, n$. 

For each model $j = 1, \ldots, J$ we compute $\PC^{(\hsp j)}$ in two simple steps.  In a first step, we find the EasyUQ solution $\hat{F}_1^{(\hsp j)}, \ldots, \hat{F}_n^{(\hsp j)}$ for the data 
\begin{align*}
\left( x_1^{(\hsp j)}, y_1 \right), \ldots, \left( x_n^{(\hsp j)}, y_n \right) ,
\end{align*}
which requires roughly ${\cal O}(n \log n)$ operations using variants of the pool-adjacent-violators (PAV) algorithm \citep{Henzi2022}.  In the second and final step, we compute
\begin{align}  \label{eq:PC}
\PC^{(\hsp j)} = \frac{1}{n} \sum_{i=1}^n \CRPS \left( \hat{F}_i^{(\hsp j)}, y_i \right) .
\end{align} 
While $\PC^{(\hsp j)}$ is the target criterion in the constrained minimization problem that defines the EasyUQ solution, the PAV algorithm does not involve the computation of the target criterion, and so this second step is mandatory, at further computational cost of up to ${\cal O}(n^2)$ operations.

In words, $\PC$ equals the mean CRPS of the EasyUQ solution when computed in retrospect from the evaluation data.  Evidently, an EasyUQ solution could also be computed from past training data, if available, and could then be used in real time to obtain probabilistic forecasts, as implemented by \citet{Walz2024b}, \citet{Lipiecki2024}, and \citet{Bulte2025}, among others.  The $\PC$ measure can be interpreted as a slightly optimistic proxy for the mean CRPS that would be obtained in such a way.  However, one could also argue that $\PC$ is a pessimistic proxy for the predictive performance of operational probabilistic products, which tend to be more sophisticated than EasyUQ.  One might then hope that the two effects balance each other, to the effect that $\PC$ is a reliable proxy for the mean CRPS of an operational probabilistic product.  We addresss and confirm this line of reasoning in the subsequent case study of WeatherBench 2 data. 

Well-known properties of the CRPS imply that the $\PC$ measure is nonnegative, is negatively oriented --- the smaller, the better --- and is reported in the unit of the outcome \citep{Gneiting2007a}.  In many settings, it is desirable to convert to a positively oriented skill score.  This can be done in the usual way, by comparing to the mean CRPS, denoted $\PC^{(0)}$, for (in-sample, unconditional) probabilistic climatology, that is, the empirical distribution of the outcomes $y_1, \ldots, y_n$.  From the representation \eqref{eq:CRPS2} it is readily seen that 
\begin{align}  \label{eq:PC0}
\PC^{(0)} = \frac{1}{2n^2} \sum_{i=1}^n \sum_{j=1}^n |y_i - y_j|
\end{align} 
equals one half times the Gini mean difference \citep{Yitzhaki2003} of the outcome variable, and we note that $\PC^{(0)}$ can be computed from the order statistics of $y_1, \ldots, y_n$ in ${\cal O}(n)$ further operations \citep[eq.~(8)]{Glasser1962}.  The $\PC$ skill ($\PCS$) measure for model $j$ relative to probabilistic climatology is then defined as  
\begin{align}  \label{eq:PCS}
\PCS^{(\hsp j)} = \frac{\PC^{(0)} - \PC^{(\hsp j)}}{\PC^{(0)}}. 
\end{align} 
The $\PCS$ measure conveniently normalizes $\PC$, as $\PCS$ has a lower bound of 0 and an upper bound of 1 (see properties (a) and (d) below), thus facilitating model comparisons.  In predictability studies, however, the use of unconditional (overall) probabilistic climatology in the reference standard $\PC^{(0)}$ in \eqref{eq:PCS} may create spurious skill due to spatial and/or temporal pooling over subsets with distinct probabilistic climatologies \citep{Hamill2006}.  To avoid displays of fictitious skill, it is then advisable to compute the reference quantity $\PC^{(0)}$ on the basis of more specific, spatially and/or temporally varying versions of climatology, if available, and we give an example of how this can be done in Figure \ref{fig:B} below.

\subsection{Properties of the PC measure}  \label{sec:properties}

In the subsequent summary of key properties of the $\PC$ measure, we drop superscripts and write $\PC$ in lieu of $\PC^{(\hsp j)}$.  Furthermore, we express the properties in the generic notation of the data at \eqref{eq:data}, with $x_1, \ldots, x_n$ being the model output and $y_1, \ldots, y_n$ being the outcomes at hand.
\begin{itemize}
\item[(a)] Interpretability: $\PC$ is negatively oriented, nonnegative, and reported in the unit of the outcome. 
\item[(b)] Perfect monotone predictor property: $\PC = 0$ if, and only if, there is a nondecreasing function $f$ such that $y_i = f(x_i)$ for $i = 1, \ldots, n$. 
\item[(c)] Invariance under strictly increasing transformations of the model output: If $g$ is a strictly increasing function, then the data $(x_1,y_1), \ldots, (x_n,y_n)$ and $(\hsp g(x_1),y_1), \ldots, (\hsp g(x_n),y_n)$, respectively, yield the same value of the $\PC$ measure. 
\item[(d)] Upper bound: $\PC \leq \PC^{(0)}$, where $\PC^{(0)}$ is the CRPS at \eqref{eq:PC0} for (in-sample, unconditional) probabilistic climatology. 
\end{itemize}

We comment on and illustrate these desirable properties within the simulation example below.  Property (a) is an immediate consequence of the definition of the CRPS \citep{Gneiting2007a}; properties (b), (c), and (d) follow readily from characteristics of IDR and EasyUQ \citep{Henzi2021, Walz2024a}.  For example, property (c) stems from the fact that the EasyUQ solution is invariant under strictly increasing transformations of the model output $x_1, \dots, x_n$.

The properties translate naturally to the skill measure $\PCS$.  Specifically, $\PCS$ is positively oriented with $0 \leq \PCS \leq 1$, $\PCS$ attains its optimal value of 1 under a perfect predictor, and $\PCS$ is invariant under strictly increasing transformations of the model output.  Finally, we emphasize that $\PC$ and $\PCS$ are measures of potential predictive ability.  They are designed to afford comparisons of single-valued model output in settings where the otherwise preferred procedure, namely, the pre-specification of a loss function or target functional \citep{Murphy1985, Gneiting2011a}, places competing models on unequal footings.

\subsection{Simulation example}  \label{sec:simulation}

For illustration, we return to the simple simulation setting of \citet{Henzi2021} and \citet{Walz2024a}.  Specifically, suppose that the weather quantity $Y$ depends on a weather state $W$, which is uniformly distributed between 0 and 10, in such a way that 
\begin{align*}
Y \mid W \sim \textrm{Gamma}(\textrm{shape} = \sqrt{W}, \, \textrm{scale} = \min \{ \max \{ W, 1 \}, 6 \}).  
\end{align*}
In words, given the weather state $W$, the outcome $Y$ follows a Gamma distribution with parameters that are functions of $W$.  The shape parameter is $\sqrt{W}$ and the scale parameter is $W$, except that we set the scale parameter to 1 if $W \leq 1$ and set it to 6 if $W \geq 6$.

We draw a sample $(w_1,y_1), \ldots, (w_n,y_n)$ of $n = 10,000$ pairs of weather states and weather outcomes.  Then we create four distinct series of deterministic model output, namely,    
\begin{align*}
x_i^{(1)} & = w_i, \\
x_i^{(2)} & = \textrm{mean} \! \left( Y \mid W = w_i \right) \! , \\
x_i^{(3)} & = \textrm{median} \! \left( Y \mid W = w_i \right) \! , \\
x_i^{(4)} & = q_{0.90} \! \left( Y \mid W = w_i \right)
\end{align*}
for $i = 1, \ldots, n$, where $q_{0.90}$ is the quantile at level 0.90 and the superscript denotes the assumed model.  The four series are strictly increasing transformations of each other and share the same information contents.

\begin{table}[t]

\caption{Evaluation results in the simulation study, with any unique best value in a column highlighted in bold.  \label{tab:sim1}}
\centering
\begin{tabular}{l|cccc|ccc}
\toprule 
Model Output & RMSE & MAE & QL$_{0.90}$ & $\PC$ & ACC & CPA & $\PCS$ \\
\midrule
$j = 1$              & 10.00         & 6.21          & 5.27          & 3.52 & .641          & .878 & .324 \\
$j = 2$ (mean)       & \textbf{7.58} & 5.09          & 2.58          & 3.52 & \textbf{.648} & .878 & .324 \\
$j = 3$ (median)     & 7.75          & \textbf{5.00} & 3.08          & 3.52 & .648          & .878 & .324 \\
$j = 4$ ($q_{0.90}$) & 12.46         & 9.79          & \textbf{1.43} & 3.52 & .647          & .878 & .324 \\
\bottomrule
\end{tabular}

\bigskip

\caption{As Table \ref{tab:sim1}, with the same model output but now the outcome squared.  \label{tab:sim2}}
\centering
\begin{tabular}{l|cccc|ccc}
\toprule 
Model Output & RMSE & MAE & QL$_{0.90}$ & $\PC$  & ACC & CPA & $\PCS$ \\
\midrule
$j = 1$              & 442          & 202          & 182          & 118  & .434          & .878  & .212 \\
$j = 2$ (mean)       & 438          & 198          & 178          & 118  & .433          & .878  & .212 \\
$j = 3$ (median)     & 439          & 199          & 179          & 118  & \textbf{.434} & .878  & .212 \\
$j = 4$ ($q_{0.90}$) & \textbf{432} & \textbf{191} & \textbf{171} & 118  & .428          & .878  & .212 \\
\bottomrule
\end{tabular}

\end{table}

In Table \ref{tab:sim1} we assess the predictive performance of the models in terms of various measures.  In addition to $\PC$, we consider root mean squared error (RMSE), the mean absolute error (MAE), and the mean piecewise linear quantile loss (QL$_{0.90}$) at level 0.90, which are classical measures of actual predictive performance for deterministic forecasts \citep{Gneiting2011a}.  Like $\PC$, they are negatively oriented with an optimal value of 0, and they are reported in the unit of the outcome.  We furthermore show two measures of potential predictive ability, namely, the classical anomaly correlation coefficient (ACC), and the coefficient of predictive ability (CPA) devised by \citet{Gneiting2022b}.  Like $\PCS$, these are dimensionless, positively oriented quantities with an optimal value of 1.  

We see that the conditional mean forecast (model $j = 2$) is best in terms of RMSE, the conditional median forecast ($\hsp j = 3)$ is best in terms of MAE, and the conditional quantile forecast ($\hsp j = 4$) is best in terms of QL$_{0.90}$.  These results can readily be explained theoretically, as squared error is a consistent scoring function for the mean functional, absolute error is consistent for the median functional, and the piecewise linear quantile loss at level $\alpha$ is consistent for the $\alpha$-quantile functional \citep{Gneiting2011a}.  The conditional median forecast ($\hsp j = 3$) also is best in terms of ACC, though without a theoretical justification being available.  The $\PC$, $\PCS$, and CPA measures are invariant under strictly increasing transformations of the model output, and so their values are the same for the four models.

In Table \ref{tab:sim2} we repeat the simulation experiment, but the model output is now used to predict the squared outcome $y_i^2$.  The conditional quantile forecast ($\hsp j = 4$) now is best in terms of RMSE, MAE, and QL$_{0.90}$, simply because it tends to be larger than its competitors.  The conditional median forecast ($\hsp j = 3$) is best in terms of ACC, again without a persuasive justification being available.  The CPA measure is invariant under strictly increasing transformations of both the model output and the outcome, so its value is unchanged from Table \ref{tab:sim1}.  In contrast, while $\PC$ and $\PCS$ are invariant under strictly increasing transformations of the model output, their values adapt to transformations of the outcome, reflecting the heavy right tail of the squared outcome by a much higher value of $\PC$ and a reduction in $\PCS$.

\section{Application to WeatherBench data}  \label{sec:WeatherBench}

We now apply the $\PC$ and $\PCS$ measures to the widely studied WeatherBench 1 \citep{Rasp2020} and WeatherBench 2 \citep{Rasp2024} datasets, aiming to assess and compare state of the art AIWP and NWP models as of 2020 and 2024, respectively. 

\subsection{In WeatherBench 1, NWP models dominate AIWP models}  \label{sec:WeatherBench1} 

\begin{figure}[t]
\centering
\includegraphics[width = \textwidth]{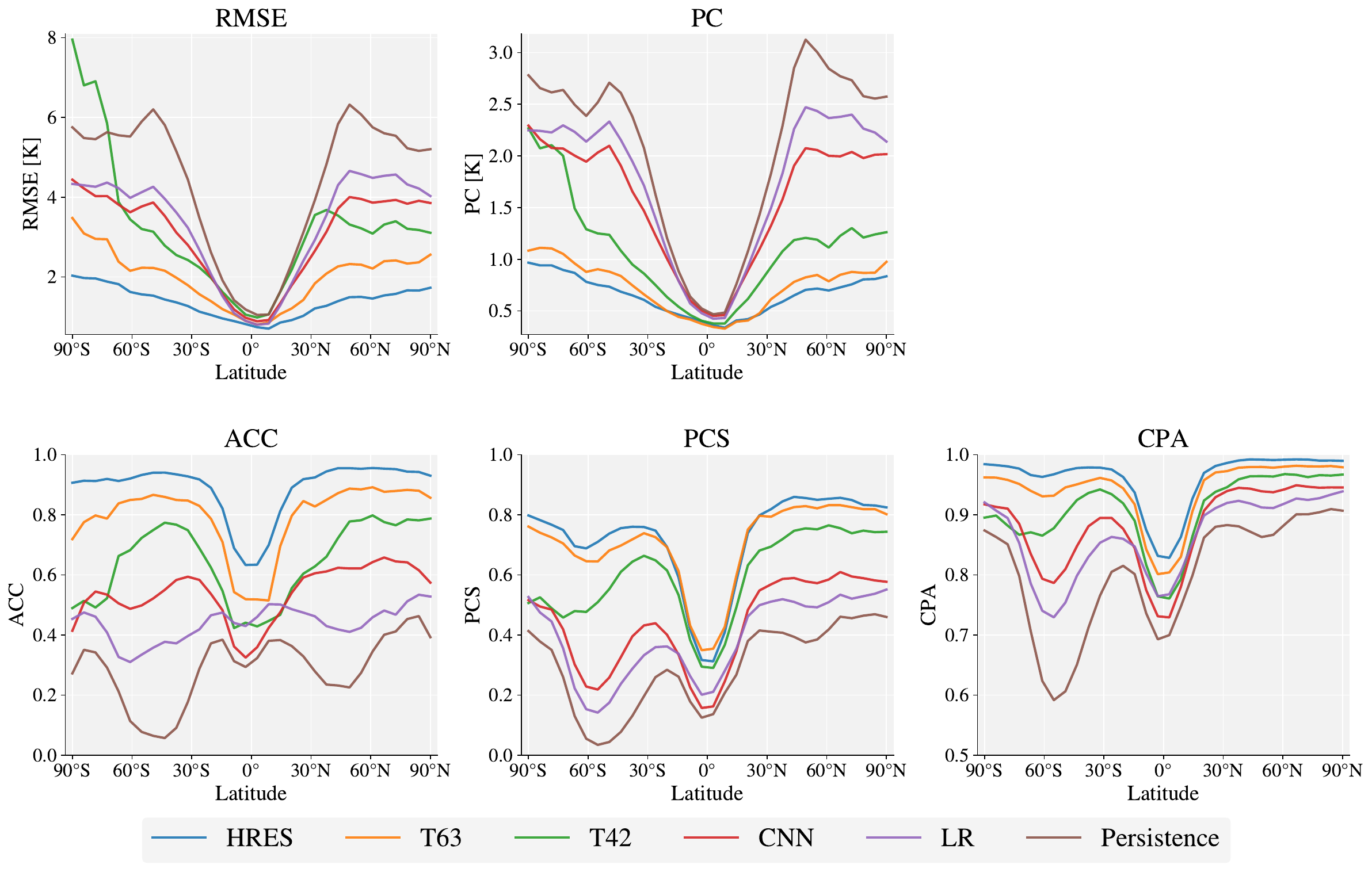}
\caption{Predictive ability of WeatherBench 1 \citep{Rasp2020} models three days ahead for 850 hPa temperature in 2017 and 2018 as a function of latitude in terms of RMSE, $\PC$, ACC, $\PCS$, and CPA, respectively.  The acronyms HRES, T63, and T42 represent physics-based NWP models run at decreasing grid resolution, which are compared to purely data-driven, neural network based (CNN), linear regression based (LR), and persistence forecasts.  The RMSE and $\PC$ measures are negatively oriented (smaller is better) and in the unit of kelvin, whereas ACC, CPA, and $\PCS$ are positively oriented and dimensionless.  The panels for RMSE and CPA and fragments of text in this legend are reproduced from \citet[Figure 9]{Gneiting2022b}.  \label{fig:WeatherBench1}}
\end{figure}

In a pioneering paper, \citet{Rasp2020} introduced WeatherBench 1 as a benchmark dataset for the comparison of AIWP model output, NWP model output, and baseline methods, such as persistence and climatology, at upper air levels, with the RMSE being the key evaluation measure.  We follow \citet{Gneiting2022b} and compare WeatherBench 1 models at a prediction horizon of three days ahead for temperature at the 850 hPa pressure level.  As performance measures we consider RMSE and $\PC$ (in kelvin) and the dimensionless ACC, CPA, and $\PCS$ measures.  The comparison includes data-driven models (CNN, LR), physics-based models (HRES, T63, T42), and persistence, and we refer to Table 2 of \citet{Rasp2020} and the legend of our Figure \ref{fig:WeatherBench1} for details.  The panels in the figure show the performance measures for the WeatherBench 1 final evaluation period as functions of latitude bands.  We compute the measures grid point by grid point, and then average across latitude bands.  While differing in detail from the assessment in terms of RMSE, ACC, and CPA, the results under the $\PC$ and $\PCS$ measures in panels (b) and (d) confirm the unequivocal superiority of the physics-based models in WeatherBench 1.

\subsection{In WeatherBench 2, GraphCast and Pangu-Weather outperform the IFS HRES model}  \label{sec:WeatherBench2}

The period since early 2022 has seen revolutionary progress in weather prediction, with data-driven AIWP models now rivaling, if not surpassing, the predictive abilities of NWP models \citep{BenBouallegue2024}.  In view of these developments, \citet{Rasp2024} devised WeatherBench 2 as a far-reaching update to WeatherBench 1.  WeatherBench 2 includes an open-source evaluation framework and makes training, ground truth, and baseline data publicly available via a continuously updated website (\url{https://sites.research.google/weatherbench}).  Retrospective forecasts at 1.5 degrees resolution are available for calendar year 2020, which serves as the WeatherBench 2 evaluation time period, and in contrast to its predecessor, WeatherBench 2 now includes forecasts for surface variables.  Following \citet{Olivetti2024}, we compare the leading data-driven GraphCast \citep{Lam2023} and Pangu-Weather \citep{Bi2023} models to the physics-based HRES model, which is the deterministic backbone of the European Centre for Medium-Range Weather Forecasts (ECMWF) ensemble \citep{Molteni1996} and ECMWF Integrated Forecast System (IFS).  We consider WeatherBench 2 forecasts of the surface weather variables mean sea level pressure (MSLP) in the unit of Pascal, 2 m temperature (T2M) in kelvin, and 10 m wind speed (WS10) in meters per second.  The forecasts are initialized at 00 UTC and 12 UTC at a spatial resolution of 1.5 degrees and have lead times of one, three, five, seven, and ten days, respectively. 

An important consideration in WeatherBench 2 is that some models are initialized and trained with, and evaluated against, the ERA5 analysis \citep{Hersbach2020} as ground truth, whereas others are initialized and fine-tuned with, and evaluated against, an ECMWF Integrated Forecast System (IFS) analysis product \citep{Rasp2024}.  This raises the question whether direct comparisons against differing types of ground thruth are meaningful.  However, for a number of AIWP models WeatherBench 2 includes distinct runs.  Specifically, the GraphCast and Pangu-Weather runs are initialized and trained with, and evaluated against, the ERA5 analysis, which is not available in real-time, whereas the HRES, GraphCast (oper.)~and Pangu-Weather (oper.)~runs in WeatherBench 2 are initialized and fine-tuned with, and evaluated against, the operational IFS analysis.\footnote{Details are provided within the GraphCast documentation at GitHub, \url{https://github.com/google-deepmind/graphcast/blob/7077d40a36db6541e3ed72ccaed1c0d202fa6014/graphcast_demo.ipynb}.}  In the subsequent comparison, we restrict attention to these five competitors, and we use the acronyms GC and PW for GraphCast \citep{Lam2023} and Pangu-Weather \citep{Bi2023}, respectively.  For concise identification, we refer to the ERA5 based runs as GC-ERA5 and PW-ERA5, and we call the IFS analysis based runs HRES, GC-IFS, and PW-IFS, respectively.

\begin{figure}[t]
\centering
\includegraphics[width = \textwidth]{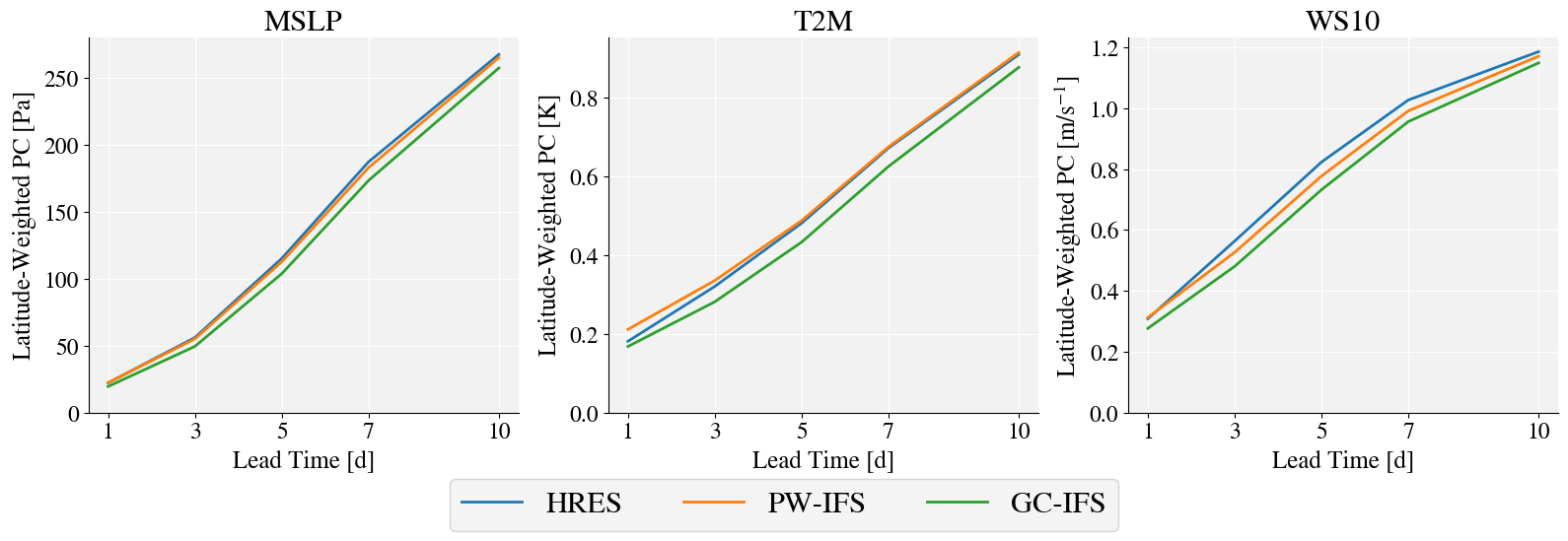}
\caption{Potential predictive ability of HRES, GC-IFS, and PW-IFS model output from WeatherBench 2 \citep{Rasp2024} for mean sea level pressure (MSLP), 2 m temperature (T2M), and 10 m wind speed (WS10) at lead times of one, three, five, seven, and ten days in terms of latitude-weighted $\PC$, with the IFS analysis serving as ground truth.  \label{fig:A}}
\end{figure}  

The cleanest and most straightforward way to proceed is to restrict direct comparisons to models that use the same analysis product as initial conditions, and to use this product as ground truth as well, as argued and implemented by \citet{BenBouallegue2024}.  Accordingly, Figure \ref{fig:A} compares HRES, GC-IFS, and PW-IFS in terms of the $\PC$ measure for the three surface weather variables, with the IFS analysis serving as ground truth.  For each lead time, we compute the $\PC$ measure grid point by grid point and find the latitude-weighted average \citep[Section 4.3]{Rasp2024}.  We note tremendously impressive performance of all three models for all three variables and all prediction horizons considered.  As noted, $\PC$ can be interpreted as a proxy for the mean CRPS of operational probabilistic products based on the deterministic backbone at hand.  For T2M, this proxy ranges from about 0.20 kelvin at a lead time of a single day to less than one kelvin ten days ahead.  For all combinations of weather variable and lead time, GC-IFS dominates both PW-IFS and HRES.  For MSLP and WS10, PW-IFS generally outperforms HRES, whereas for T2M, HRES outperforms PW-IFS.

\begin{figure}[t]
\centering
\includegraphics[width = \textwidth]{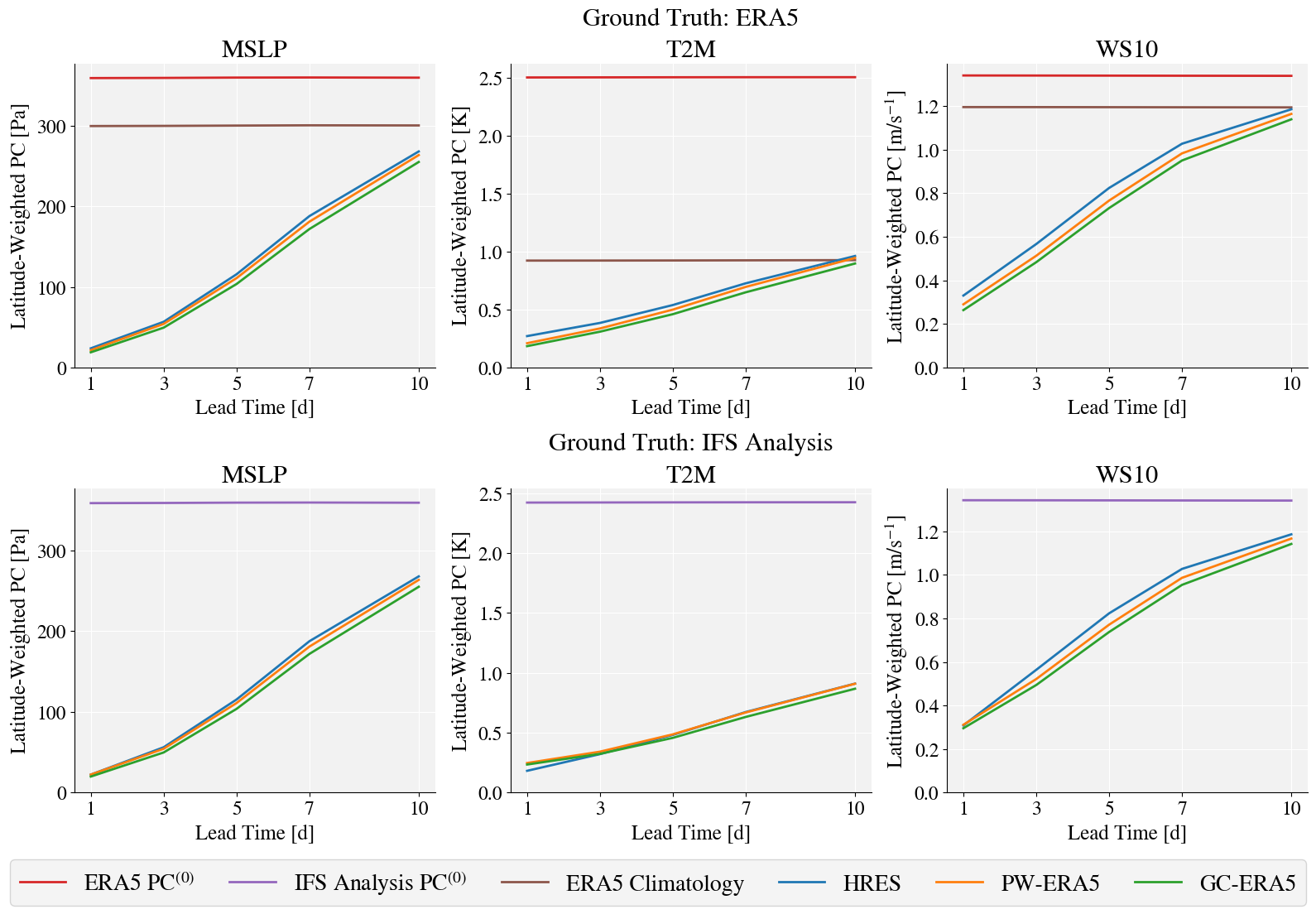}
\caption{Potential predictive ability of HRES, GC-ERA5, and PW-ERA5 model output from WeatherBench 2 \citep{Rasp2024} for mean sea level pressure (MSLP), 2 m temperature (T2M), and 10 m wind speed (WS10) at lead times of one, three, five, seven, and ten days in terms of latitude-weighted $\PC$, with either ERA5 (top panel) or the IFS analysis (bottom panel) serving as ground truth.  For reference, we also show the latitude-weighted $\CRPS$, denoted $\PC^{(0)}$, for (unconditional) probabilistic climatology based on the ERA5 analysis and the IFS analysis, respectively, and latitude-weighted $\PC$ for the seasonally varying ERA5 climatology forecast from WeatherBench 2.  \label{fig:C}}
\end{figure}

In practice, there might be settings where a direct comparison is required, even though the models are initialized with distinct analyses.  For illustration, we compare the HRES model to the GC-ERA5 and PW-ERA5 models.  A natural approach is to compute the $\PC$ measure twice, once with the IFS analysis serving as ground truth and once more with the ERA5 analysis as ground truth.  If the respective rankings disagree, we might take a closer look and apply resampling-based tests to check for statistical significance.  Figure \ref{fig:C} shows the latitude-weighted $\PC$ measure for HRES, GC-ERA5, and PC-ERA5 with ground truth being either the ERA5 analysis (top row) or the IFS analysis (bottom row).  As baseline, we also find the latitude-weighted $\CRPS$, denoted $\PC^{(0)}$, for (unconditional) probabilistic climatology based on the ERA5 analysis and the IFS analysis, respectively.  As noted in Section \ref{sec:properties}, unconditional probabilistic climatology is a rather weak competitor, due to the pooling over seasonally varying probabilistic climatologies.  In this light, we add a further baseline to the top row panels, namely, latitude-weighted $\PC$ for the seasonally varying (deterministic) ERA5 climatology forecast from WeatherBench 2.  However, WeatherBench 2 does not include an IFS based climatology forecast, so the bottom row panels lack the further baseline.  Comparing to the stronger, more realistic baseline, we see that GC-ERA5, PW-ERA5, and HRES outperform seasonally varying climatological forecasts up to, and including, a lead time of ten days for MSL, whereas for T2M and WS10, current predictability limits are at about ten days.  For MSLP and WS10, the rank order of GC-ERA5, PW-ERA5, and HRES is generally the same under either the ERA5 analysis or the IFS analysis serving as ground truth, and the order agrees with the rank order of GC-IFS, PW-IFS, and HRES in Figure \ref{fig:A}.  For T2M, however, the rank order in Figure \ref{fig:C} depends on the type of analysis that serves as ground truth.

\begin{figure}[t]
\centering
\includegraphics[width = \textwidth]{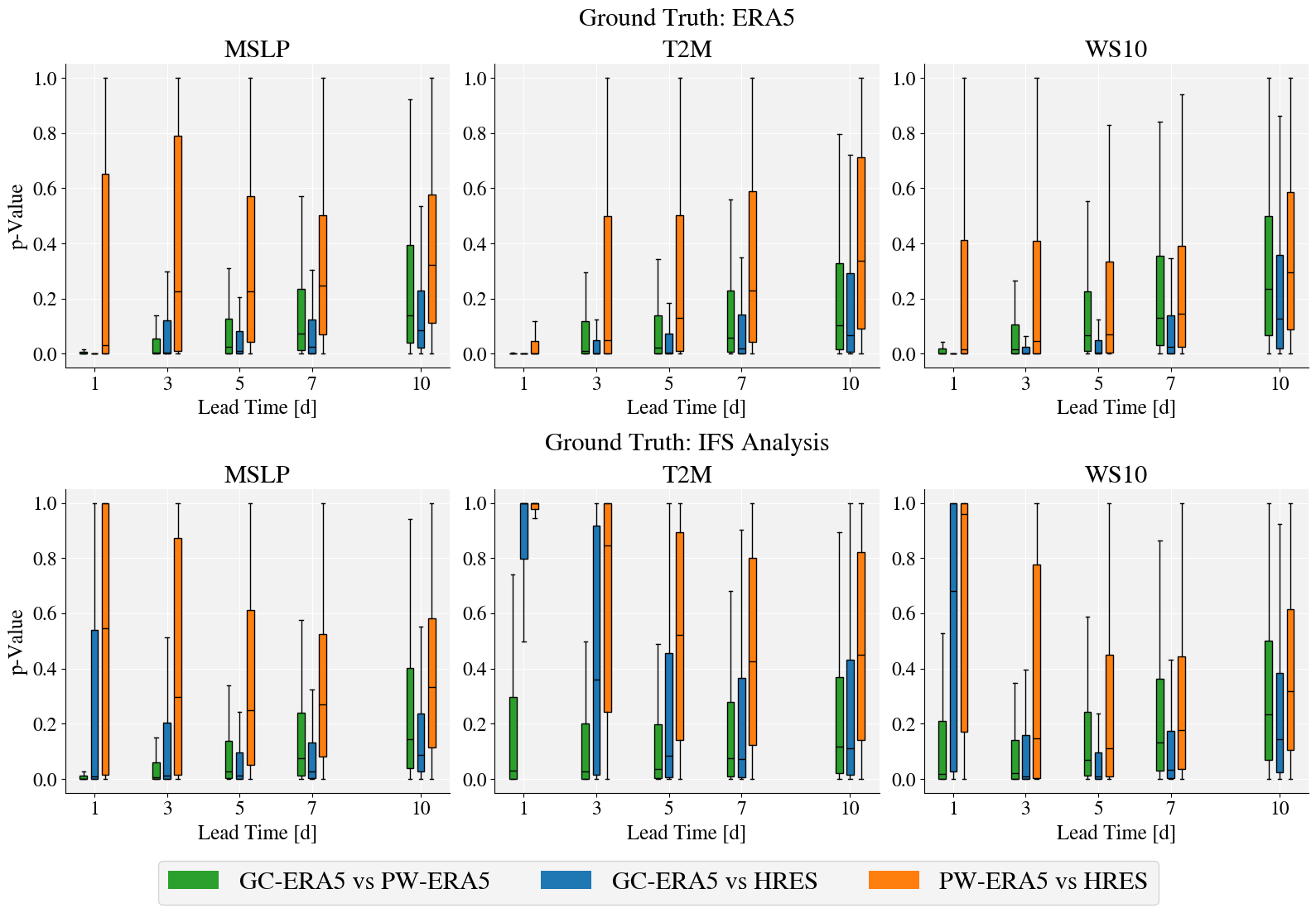}
\caption{Statistical significance of results from Figure \ref{fig:C}.  The boxplots summarize one-sided $p$-values at the grid point level under the null hypothesis of model A and model B having the same $\PC$.  Small $p$-values suggest rejection in favour of model A (first named); large $p$-values suggest rejection in favour of model B (second named).  \label{fig:D}}
\end{figure}  

The ambiguity in the rank order for T2M suggests that we assess the statistical significance of differences in $\PC$, for which we propose the use of permutation tests.  \citet{Feldmann2019} devised block bootstrap techniques to generate confidence bands in related settings, and we adapt their methods to a block permutation technique for testing the (null) hypothesis of equal predictive performance in terms of expected $\PC$.\footnote{WeatherBench 2 comprises two daily runs initialized at 00 UTC and 12 UTC, respectively.  Therefore, successive forecasts and outcomes at a lead time of $k$ days must be assumed to be dependent at lags less than $2k$.  In this light, when comparing model A and model B at a lead time of $k$ days, we consider successive blocks of length $2k$ of the respective score difference.  We assign a random sign to each block and concatenate these blocks, to form a block permutation sample of the same size ($n = 366 \times 2 = 732$) as the actual series.  Then we find the rank $R$ of the actual difference in $\PC$ when pooled with the differences from $N = 1,000$ block permutation samples, and use $p = R/N$ as a one-sided $p$-value under the null hypothesis of equal predictive performance.}  Our conventions are such that, when comparing model A and model B, $p$-values close to 0 suggest rejection in favour of model A outperforming model B, whereas $p$-values close to 1 suggest vice versa.  In Figure \ref{fig:D} we show boxplots of the $p$-values at the $240 \times 121 = 29,040$ grid points at a spatial resolution of 1.5 degrees.  For MSLP and WS10, the $p$-values support the rank order from Figures \ref{fig:A} and \ref{fig:C}, in that GraphCast outperforms Pangu-Weather, and Pangu-Weather outperforms HRES.  For T2M, GC-ERA5 outperforms PW-ERA5.  However, the rankings that involve HRES depend on the type of ground truth and/or the lead time, and at very many grid points the respective difference in the PC measure fails to be statistically significant.  We conclude that for T2M the differences in the predictive performance of the deterministic backbone of GraphCast, Pangu-Weather, and HRES, respectively, which are evident from Figure \ref{fig:A}, get masked under the confounding effects of distinct initial conditions.

\begin{figure}[t]
\centering
\includegraphics[width = 0.70 \textwidth]{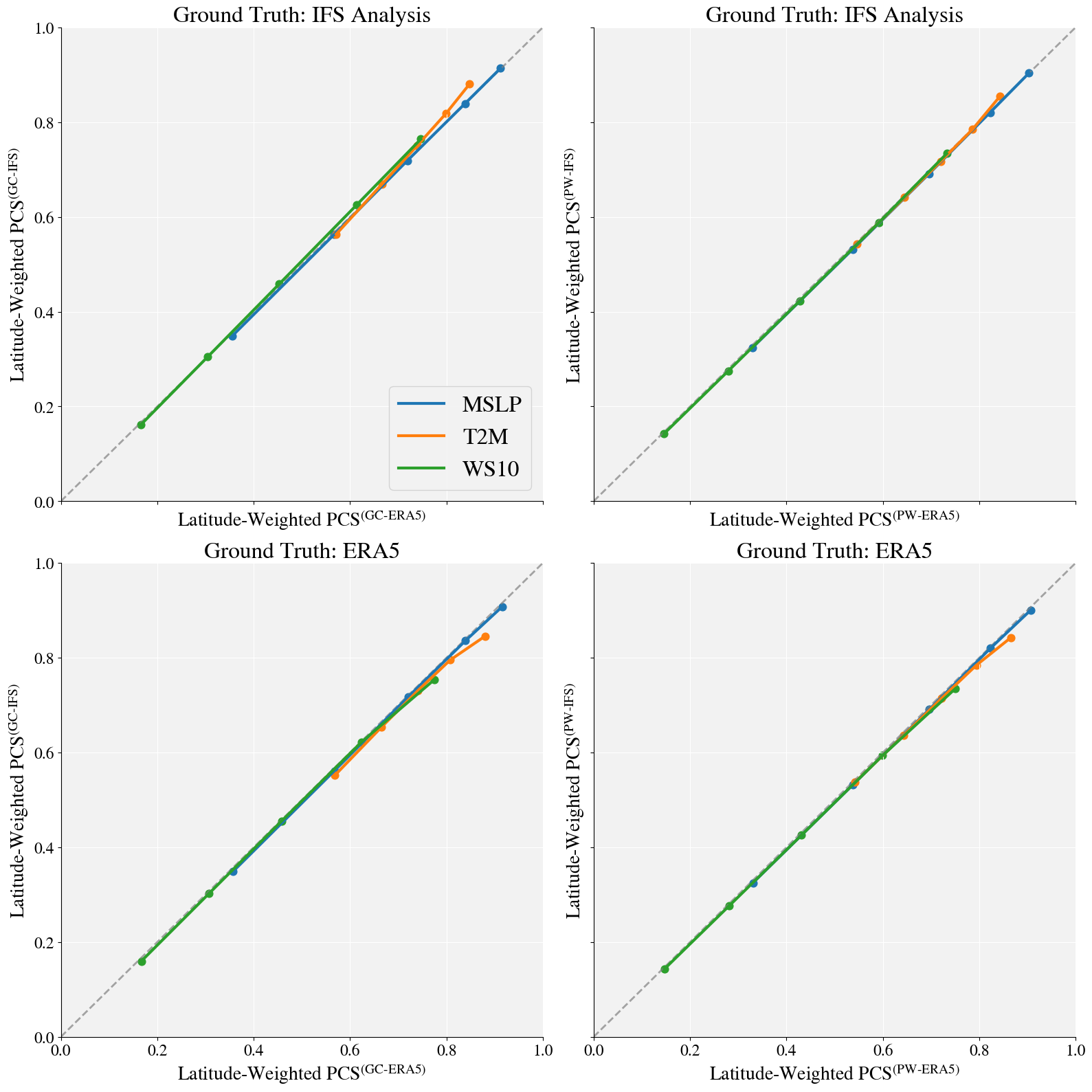}
\caption{Scatterplot of latitude-weighted $\PCS$ of (left column) GC-ERA5 versus GC-IFS and (right column) PW-ERA5 versus PW-IFS with (upper row) the IFS analysis and (lower row) the ERA5 analysis serving as ground truth.  For each variable, the points represent lead times of one, three, five, seven, and ten days, respectively.  \label{fig:E}}
\end{figure}  

Generally, if we use the ERA5 analysis as ground truth, and compare any given model initialized under either the ERA5 or the IFS analysis, the former ought to have a competitive advantage.  In contrast, if  we use the IFS analysis as ground truth, it is not clear which of the two approaches is superior.  On the one hand, a model version initialized with the IFS analysis has the advantage that its initial conditions correspond to the ground truth.  On the other hand, the ERA5 analysis is based on an assimilation window that extends hours into the future \citep{Hersbach2020}, and thus it may provide a competitive advantage relative to the IFS analysis, which is available in real time, particularly for T2M.  We study these relationships in Figure \ref{fig:E}, where we show scatterplots of latitude-weighted $\PCS$ for the IFS-initialized versus the ERA5-initialized runs of GraphCast and Pangu-Weather, respectively.\footnote{We compute the latitude-weighted $\PCS$ measure in the same way as \eqref{eq:PCS}, but with each term on the right-hand side replaced by the respective latitude-weighted sum of $\PC$ values.}  If we use the ERA5 analysis as ground truth (lower row), then indeed the ERA5 initialized versions perform better, with the effect being particularly pronounced for T2M and GraphCast at a lead time of a single day.  In contrast, if we use the IFS analysis as ground truth, there is no consistent pattern of either the ERA5 analysis or the IFS analysis initialized versions being superior.

When considering the AIWP models, one might also wish to compare $\PCS$ for GC-IFS under the IFS analysis as ground truth to $\PCS$ for GC-ERA5 under the ERA5 analysis as ground truth, and similarly for Pangu-Weather.  The largest difference between the respective scores across the $2 \times 3 \times 5 = 30$ combinations of AIWP model (GraphCast, Pangu-Weather), weather variable, and lead time considered here, arises for the T2M forecast of PW-IFS and PW-ERA5 at a lead time of one day, where the skill is~.843 under the IFS analysis and~.865 under ERA5.  For the 29 other combinations, none of the differences exceeds~.018.  The superb alignment testifies to the quality and adequacy of both the $\PCS$ measure and the two types of ground truth data in WeatherBench 2. 

\begin{figure}[tp]
\centering
\includegraphics[width = \textwidth]{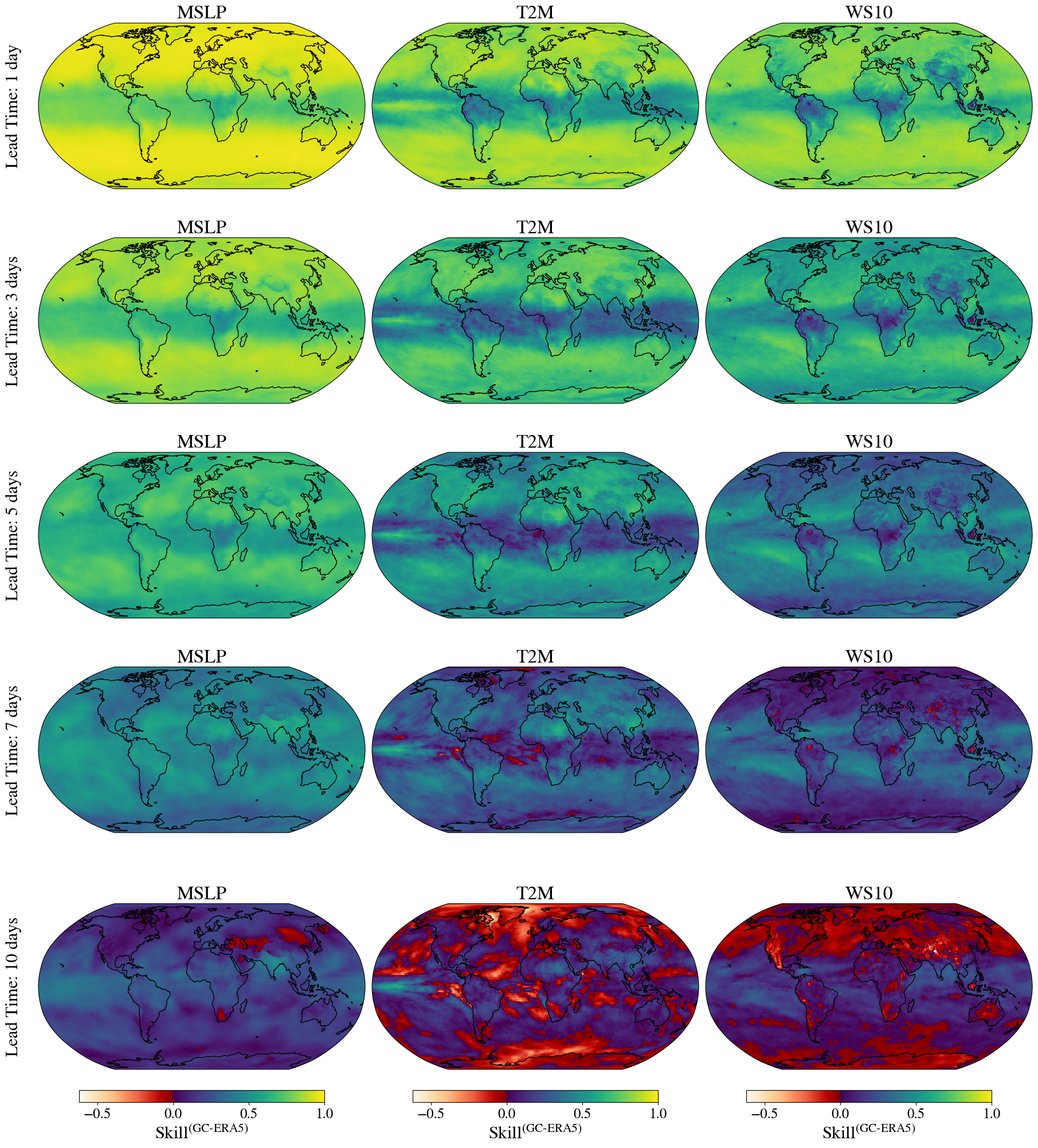}
\caption{Grid point level potential predictive ability of GC-ERA5 for mean sea level pressure (MSLP), 2 m temperature (T2M), and 10 m wind speed (WS10) at lead times of one, three, five, seven, and ten days in terms of $\PC$ skill relative to the ERA5 climatology forecast from WeatherBench 2, with the ERA5 analysis serving as ground truth.  \label{fig:B}}
\end{figure}  

Finally, to showcase spatial patterns in potential predictive ability, Figure \ref{fig:B} displays the $\PC$ skill of the GC-ERA5 model relative to the ERA5 climatology forecast from WeatherBench 2.\footnote{The patterns for the other models are very similar, though generally with a more rapid decay of the skill.}  While the top panel of Figure \ref{fig:C} shows latitude-weighted $\PC$ skill of the ERA5 climatology forecast, we now turn to the grid point level.  The panels display skill relative to a seasonally varying climatology, so as lead time increases we expect the skill to decay to 0, and eventually become negative, for all variables and at all locations.  For T2M and WS10 the predictability limit is reached at lead times of seven to ten days, whereas for MSLP it may extend beyond ten days, particularly in the tropics.  We generally note decreased skill over the tropics when compared to the middle latitudes at shorter lead times, but the relationship appears to reverse at larger lead times.  The differential skill is broadly consistent with recent findings by \citet{Keane2025}, who note a ``crossover'' in performance between tropical and mid latitudes at lead times of five to seven days for both AIWP and NWP models.

\subsection{The PC measure of the HRES model serves as proxy for the mean CRPS of the operational ECMWF ensemble}  \label{sec:IFS}

Following \citet{Brenowitz2025}, a natural question is whether $\PC$ could be used as a proxy for the performance of an operational ensemble system constructed from the deterministic backbone at hand, as measured by the mean CRPS.  On the one hand, $\PC$ appears to be an optimistic proxy, because EasyUQ is based on an in-sample optimization of the CRPS.  However, one might also argue that $\PC$ is a pessimistic proxy, as an operational ensemble is bound to use more information than is contained in the training data for EasyUQ.  Heuristically, the additional information (e.g., pointing at current and future weather regimes) that an operational ensemble has available, becomes more and more critical as lead time increases.  Thus, one might assume that, while the two effects generally balance each other, the proxy might tend to be optimistic at short lead times and pessimistic at larger lead times.  

\begin{figure}[tp]
\centering
\includegraphics[width = \textwidth]{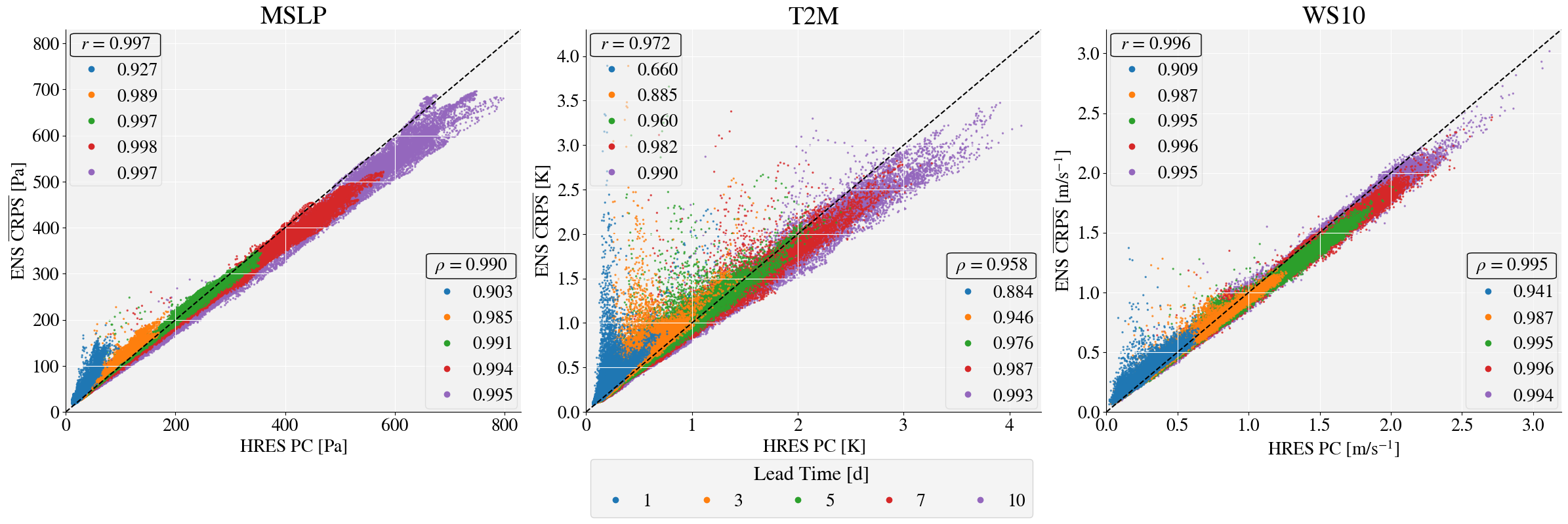}
\caption{Scatterplots of the mean CRPS for the operational ECMWF ensemble versus $\PC$ for IFS HRES model output, for mean sea level pressure (MSLP), 2 m temperature (T2M), and 10 m wind speed (WS10) at lead times of one, three, five, seven, and ten days, with the IFS analysis serving as ground truth.  In each panel, we also show the Pearson correlation coefficient ($r$) and Spearman's rank correlation coefficient ($\hsp \rho$) at individual lead times and aggregated across lead times.  \label{fig:WeatherBench2d}}
\end{figure}  

To investigate this hypothesis, Figure \ref{fig:WeatherBench2d} shows scatterplots of the mean CRPS of the operational ECMWF ensemble against $\PC$ for the IFS HRES model, for mean sea level pressure (MSLP), 2 m temperature (T2M), and 10 m wind speed (WS10), respectively.  We compare the operational CRPS and the $\PC$ measure grid point by grid point and at lead times of one, three, five, seven, and ten days, with the IFS analysis serving as ground truth.  For all three variables, $\PC$ is a meaningful proxy for the mean CRPS of the operational ensemble, with the proxy being somewhat optimistic at the shortest lead time of one day --- as indicated by a surplus of points above the diagonal --- and being pessimistic at lead times of seven and ten days.  

For comparison, the scatterplots in Figure \ref{fig:WeatherBench2d} are of the same type\footnote{Note that relative to our figure the axes in Figure 1(b) of \citet{Brenowitz2025} are interchanged, and the display is for a distinct variable.} as Figure 1(b) of \citet{Brenowitz2025}, where the CRPS of the operational ECMWF ensemble is compared to the proposed proxy, namely, the mean CRPS of a lagged ensemble that is constructed from the deterministic backbone at hand.  For our proxy, the correlations are much higher and very close to the optimal value of 1.  More importantly, the point cloud in Figure 1(b) of \citet{Brenowitz2025} is far off the diagonal, illustrating that the proposed proxy yields a decidedly optimistic estimate of the mean CRPS of the operational ensemble.  Arguably, the pronounced bias stems from the lagged ensemble having members with shorter lead times than the operational ensemble.  In contrast, our approach does not invoke forecasts at shorter lead times, and the scatterplots in Figure \ref{fig:WeatherBench2d} remain close to the diagonal.

\section{Discussion}  \label{sec:discussion} 

With the advent of data-driven weather prediction, meaningful ways of comparing the deterministic backbone of physics- and AI-based weather forecasts have been sought.  Recently, \citet{Brenowitz2025} introduced a framework based on lagged initial condition ensembles that compares data-driven and physics-based models with the same ensemble technique.  Our approach is of a related spirit, as we also build probabilistic forecasts from a deterministic forecast archive.  Specifically, we use the EasyUQ technique \citep{Walz2024a} to convert single-valued model output into an in-sample optimal probabilistic forecast.  The $\PC$ measure then equals the mean CRPS for the thus generated probabilistic forecasts; it is reported in the unit of the outcome and serves as a proxy for the mean CRPS of operational probabilistic products.  Across application disciplines, the $\PC$ and $\PCS$ measures allow for meaningful comparisons of single-valued forecasts in settings where the pre-specification of a loss function or functional --- which is the usual, and principally superior, procedure in forecast contests, and for administrative or benchmarks purposes --- places competitors on unequal footings.  Code for the computation of $\PC$ and $\PCS$ and replication material for this paper is available in \textsf{R} \citep{R} and Python \citep{Python}.\footnote{See \url{https://github.com/tobiasbiegert/potential-crps}.}

We emphasize that the $\PC$ measure is tailored to the evaluation of single-valued model output.  AI-based weather forecasts are increasingly becoming probabilistic \citep{Kochkov2024}, and in probabilistic settings the CRPS can be used as a loss function for training \citep{Lang2024}.  However, for probabilistic forecasts the alignment (or not) of the loss functions used for training and evaluation, respectively, is much less of a concern than for single-valued forecasts.  The reason for this that any strictly proper scoring rule \citep{Gneiting2007a} incentivizes the same goal which, arguably, is shared by physics-based models, namely, to maximize the sharpness of the probabilistic forecasts subject to calibration \citep{Gneiting2007b}.  In contrast, for single-valued forecasts distinct loss functions might entail vastly distinct incentives, and in the absence of a loss function specification it is not clear what directions a physics-based model might follow \citep{Murphy1985, Gneiting2011a}.

A possible limitation of our approach is that it requires evaluation sets of a size that is sufficiently large to allow for purposeful application of EasyUQ and sufficiently small to allow computations.  In the experiments on WeatherBench 1 and WeatherBench 2 data, we computed $\PC$ at individual grid points for forecasts once or twice daily over periods of one or two years, for sample sizes $n$ of about 730, where the method yields meaningful results.  Generally, our method handles evaluation sets with sizes $n$ in the hundreds to a few thousands.  For larger $n$, the method can be adapted by implementing EasyUQ via subset aggregation \citep{Henzi2021}, which at present requires user input to tune subset sampling parameters, but could be automated in future work. 

In our case study of WeatherBench 2 forecasts, we followed the lead of \citet{BenBouallegue2024} and emphasized the comparison of the IFS-initialized GraphCast, Pangu-Weather, and ECMWF HRES model output, which use the same initial conditions.  In this comparison, GraphCast outperformed both Pangu-Weather and HRES in terms of $\PC$.  Subsequent comparisons to ERA5 initialized versions showcased the confounding effects of distinct initial conditions, particularly for surface temperature at short lead times.  For the HRES model, $\PC$ emerges as a strikingly close proxy for the mean CRPS of the operational ECMWF ensemble.  We emphasize that our analysis concerns the quality of model output for univariate weather variables at individual lead times and individual locations.  While $\PC$ and $\PCS$ address and resolve major shortcomings in previously used evaluation methodology in such settings, they do not address inter-variable dependency or the physical consistency of spatio-temporal forecast trajectories \citep{Bonavita2024}.  If these more complex, multivariate matters are of prime interest, $\PC$ and $\PCS$ ought to be accompanied by comprehensive suites of evaluation tools that are tailored to them \citep{BenBouallegue2024, Radford2025b}.  

We close the paper with technically oriented thoughts on further methodological development.  Forecast evaluation with focus on extreme weather is of much current interest \citep{Lerch2017, Olivetti2024}.  To address such a focus, the universality of the IDR solution \citep[][Theorem 2]{Henzi2021} allows for the replacement of the CRPS in the target criterion at \eqref{eq:CRPS1} by a threshold- or quantile-weighted CRPS \citep{Gneiting2011b} or even a single, piecewise linear quantile loss function.  In the latter case, the EasyUQ technique could be replaced by a straightforward, single application of the pool-adjacent-violators (PAV) algorithm \citep{Jordan2022}, and the isotonicity assumption in terms of the stochastic order under EasyUQ reduces to a monotonicity assumption at the quantile level at hand.  Methods of this type are tailored to extremes, but avoid forecast evaluation conditional on extreme outcomes and thus address concerns raised by \citet{Lerch2017}.

In our case study on WeatherBench 2 data, we used a simple block permutation method to assess the statistical significance of a difference in $\PC$.  From the perspective of mathematical statistics one might, alternatively, invoke recently developed theory \citep{Arnold2025} and view the $\PC$ measure as a functional of the joint distribution of the model output and the weather outcome at a (hypothetical) population level.  A related point of view has been taken in the biostatistical literature, where researchers seek to estimate expected Brier scores at the population level \citep{Gerds2006}.  In our setting one might then leverage asymptotic distribution theory established by \citet{Mosching2020} to devise formal statistical tests for the (null) hypothesis of equal $\PC$ values at the population level.  While developments in these directions are bound to be highly complex, we encourage future work, possibly starting from the just mentioned special case of a single, piecewise linear quantile loss function in lieu of the CRPS.  Despite the formidable technical challenge, methods of these types are bound be of great interest not only in the evaluation of AIWP and NWP models, but in a gamut of forecasting problems where the pre-specification of a loss function puts competitors on unequal footings.

\section*{Acknowledgements}

We thank Peter Knippertz, Victoria Stodden, and Johanna Ziegel for fruitful discussion.  Tilmann Gneiting, Alexander Jordan, Kristof Kraus, and Eva-Maria Walz are grateful for the generous support of the Klaus Tschira Foundation.  Tobias Biegert gratefully acknowledges support by the German Weather Service (Deutscher Wetterdienst) through the SPARC-ML project within the extramural research program, funding reference number 4823EMF01.  Sebastian Lerch gratefully acknowledges support by the Vector Stiftung through the Young Investigator Group ``Artificial Intelligence for Probabilistic Weather Forecasting''.  The research leading to the results presented in Sections \ref{sec:WeatherBench2} and \ref{sec:IFS} has been supported by a Google Cloud Credit Award.

\bibliographystyle{apalike}
\bibliography{biblio}

\end{document}